\documentclass[referee]{aa} 
\usepackage[T1]{fontenc}
\usepackage{amsmath}

\usepackage{graphicx}
\usepackage{txfonts}
\usepackage{natbib} 
\bibpunct{(}{)}{;}{a}{}{,} 

\newcommand{\ve}[1]{{\bf #1}}

\begin{document}

\title{Towards a theory of extremely intermittent pulsars}
\subtitle{I: Does something orbits PSR B1931 + 24 ?}


   \author{F. Mottez \inst{1} \and S. Bonazzola \inst{1} 
          \and
          J. Heyvaerts \inst{2,1}\fnmsep
          }


   \institute{LUTH, Observatoire de Paris, CNRS, Unviersit\'e Paris Diderot,
              5 place Jules Janssen, 92190 Meudon, France.\\
              \email{fabrice.mottez@obspm.fr}
         \and
             Universit\'e de Strasbourg\\
             \email{heyvaerts@astro.u-strasbg.fr}}


 
  \abstract 
   {}
  {We investigate whether one or many companions 
    are orbiting the extremely intermittent pulsar PSR B1931+24.}
   {We constrained our analysis on previous observations of eight fundamental properties of PSR B1931+24. 
The most puzzling properties are the intermittent nature of the pulsar's activity, with active and quiet phases 
that alternate quasi-periodically; the variation of the slowing-down rate of its period between active and quiet phases; 
and because there are no timing residuals, it is highly unlikely that the pulsar has a massive companion. 
Here, we examine the effects that one putative companion immersed in the magnetospheric plasma 
or the wind of the pulsar might have, as well as the associated electric current distribution. 
We analysed several possibilities for the distance and orbit of this 
hypothetical companion and the nature of its interaction with the neutron star.}
   {We show that if the quasi-periodic behaviour of PSR B1931+24 was caused 
by a companion orbiting the star with a period of 35 or 70 days, the radio emissions, 
usually considered to be those of the pulsar would in that specific case be emitted in
the companion's environment. We analysed four possible configurations and conclude that none of them 
would explain the whole set of peculiar properties of PSR 1931+24. We furthermore considered a period 70 days for the precession of the periastron associated to an orbit very close to the neutron star. 
This  hypothesis is analysed in a companion paper.}


   \maketitle
%

\newpage
\section{Introduction}

The pulsar \object{PSR B1931+24}, qualified as extremely intermittent, is characterised by two radiation regimes. 
The on regime corresponds to the normal radiation of a standard pulsar. 
The off regime consists of interruptions in the radio emissions that can last for days or months. 
We here discuss whether there may be a planet, or a stream of smaller bodies, in orbit around this atypical pulsar. 

Only two pulsars, \object{PSR B1257+12} and \object{PSR B1620-26}, are known to host  companions.  
They were detected through the pulsar timing method \citep{Wolszczan_1992,Thorsett_1993}. 
The statistical analysis performed in \citet{rea_2008} has shown  that 
the timing residuals of PSR B1931+24 do not give any clue that there might be an Earth-like  companion, while the range of orbital inclination angle (with respect to the line of sight) that would allow a Jupiter-like  companion is narrow. 

We explore the possibility that  a  planet in orbit around PSR B1931+24, or several asteroid-like bodies that are
small  enough  to be undetected through the pulsar timing method, behave like unipolar inductors and generate  
Alfv\'en wings that could explain PSR B1931+24's singular behaviour. 

We propose new ideas concerning the nature of these transient pulsars.
This is based on simple orders-of-magnitude derivations. We do not 
claim, at this stage of our research, to provide a detailed analysis.

PSR B1931+24 has a number of peculiar properties that have been discovered, notably,
by \citet{Kramer_2006}; they are summarised below and are numbered 
because we frequently need to refer to them in the following sections.
\begin{itemize}
\item {\bf P1} With two modes of radio emissions, PSR B1931+24 does not radiate in a standard way. It behaves like an ordinary pulsar during active (on) phases, then switches off and remains undetectable (silent / off phases). 
\item {\bf P2} The active phases last for 5-10 days, the off phases for  25-35 days long. This pattern is repeated quasi-periodically. 
\item {\bf P3} During the active phases, the pulsar slows down  at a rate ($\dot{\Omega}_{on} = -10.24 \times 10^{-14}$ s$^{-2}$)  
faster than during the quiet phases 
($\dot{\Omega}_{off} = -6.78 \times 10^{-14}$  s$^{-2}$). 
\item {\bf P4} The period of PSR B1931+24 is $P=0.813$ s. In the $P/\dot P$ diagram the pulsar is placed in the standard "second pulsar" family, far from the death line. 
Therefore, the origin of PSR B1931+24's intermittency is not that the pulsar is on the verge of extinguishing.
\item {\bf P5} Because $\dot{P}_{off} < \dot{P}_{on}$ the difference between the two regimes is not a mere change of beam direction, but also a change of the torque exerted on the star. 
Considering the variations of rotational energy, \citet{Kramer_2006} have estimated that 
the additional torque $\Delta T$ during the active phase is due to an excess electric current $\Delta I_{pc}$ 
flowing in the polar cap plasma, such that
\begin{eqnarray} \label{additional_torque_kramer}
& &\Delta T \sim \frac{2}{3c} \Delta I_{pc} B_0 \, R_{pc}^2 \mbox{{with}} 
\nonumber \\
& &\Delta I_{pc} = \pi R_{pc}^2 \Delta \rho c= 8\times 10^{11} \mbox{ A}.
\end{eqnarray}
The excess charge density $\Delta \rho \sim 0.034$ C m$^{-3}$ is similar to 
the Goldreich-Julian charge density, which for a corotating plasma is
$\rho_{GJ}= -2 \varepsilon_o {\mathbf{B}}_{*} \cdot {\boldsymbol{\Omega}}$. 
The polar cap radius $R_{pc}=(2 \pi R_*^3 \Omega /c)^{1/2}$ was computed for a star radius $R_*=10$ km, 
and the dipole magnetic field is estimated from 
$B_{*}/ \mbox{Tesla} = 3.2 \times 10^{15} (-\dot \Omega_{off}/\Omega^3)^{1/2}$, i.e. 
from the well-known spin-down rate of a rotating magnetic dipole in empty space \citep{Deutsch_1955}. 
In the following estimates, we kept the same values of $R_{*}, B_{*}$ and $M_* = 1,4M_{\odot}$ as used in \citet{Kramer_2006}.
\item {\bf P6} One transition from the on to the off phase has been observed. 
It was very abrupt, occurring in fewer than ten pulsar rotations. 
{From this it can be inferred that the transitions between on and off phases \textit{may be} sudden, but
this does not imply that all transitions between on and off states are so.} 
\item {\bf P7} The period change between the on and off phases is not measurable. 
Only the time derivative of the period changes in a measurable way.
\item {\bf P8} No companion is detectable from timing residuals.
\end{itemize}

The basic data about PSR B1931+24 used throughout the paper are displayed in Table \ref{table_parametres}.

\begin{table*}
\caption{Basic estimates about the PSR B1931+24.} 
\label{table_parametres} 
\centering 
\begin{tabular}{| l |l |} 
\hline
 stellar radius &  $R_*=10$ km  \\
 stellar mass  & $M_* = 1,4$  solar mass $=2.7 \times 10^{30}$ kg \\
 stellar surface magnetic field (pole) &  $B_* =  2.6 \times 10^{8}$ T  \\
 observed period  &  $P=0.813$ s$^{-1}$ \\
 observed pulsation  &  ${\Omega}_*=7.728$ s\\
observed frequency shift when on  &  $\dot{\Omega}_{on} = -10.24 \times 10^{-14}$ s$^{-2}$ \\
observed frequency shift when off  &  $\dot{\Omega}_{off} = -6.78 \times 10^{-14}$  s$^{-2}$ \\
light cylinder radius & $R_{lc} = c/\Omega  = 3,879 \times 10^7$  m $= 2,6 \times 10^{-4}$  AU\\
polar cap radius &$R_{pc} =R_* (\Omega R /c)^{1/2}$  = 160  m\\
Goldreich-Julian current  &  $I_{GJ}=-2 c \pi R_{pc}^2 \epsilon_0 \Omega_* B_*=8.6 \times 10^{11}$ A \\
Kramer current  &  $\Delta I_{pc} = 8\times 10^{11}$ A\\
\hline 
semi-major axis for a 35-day orbit & $ a_{35} $ = $3,49597 \times 10^{10}$  m $=0.23$ AU\\
\hline 
\end{tabular}
\end{table*}


Can we explain these properties ?

We can reject the possibility that a body orbiting the neutron star induces a precession that would change the direction of the pulsar emission beam for two reasons: 
(1) the off and on phases are too irregular for a phenomenon only induced by gravitational effects, and (2) it would not explain
why PSR B1931+24 does not radiate as regularly as most pulsars of its family. This is
not caused by the pulsar itself, but by some peculiarity of its environment. 

\citet{Cordes_2008} proposed that pulsar nullings (periods without radio emission) 
are caused by in-falls of asteroids created in the supernova remnant. 
These satellites are supposed to fall as solid matter and would be evaporated in the vicinity
of the neutron star. 
Then the plasma formed from the vaporized asteroid would be gravitationnally
attracted to the polar cap of the pulsar
and would interrupt a significant fraction of the Goldreich-Julian current, as shown by \citet{Kramer_2006}. 
This model, however, does not explain very well why PSR B1931+24 would have such long and quasi-periodic nullings.  
We would instead expect a more chaotic sequence of nullings. A possible answer might be that the asteroids 
are grouped in a swarm. This would impose specific constraints 
on the distribution of these asteroid swarms. 
Why would the nullings last for days, when the nullings explained for other pulsars last only for a few seconds ?

Objects such as PSR B1931+24 and PSR J1841-0500 differ from  
rotating radio transients \citep{McLaughlin_2006}, which typically radiate for only a few milliseconds 
once every few minutes or hours. These time scales considerably differ from
those observed in PSR B1931+24 and PSR J1841-0500. According to \citet{Lyne_2009}, 
rotating radio transients could be
exhausted magnetars. Most of them occupy the region of the $P$--${\dot{P}}$ diagram  
with periods of several seconds and ${\dot{P}}$ higher than 10$^{-14}$, although one 
is surprisingly found in the region of normal radio-pulsars. The lack of quasi-periodicity
for alternating emitting and non-emitting regimes in rotating radio transients supports the view that 
their nature is different from that of PSR B1931+24 and PSR J1841-0500.
 
Because PSR J1841-0500 is less well documented than PSR 1931+24, we focused our investigation on PSR 1931+24. 
We consider PSR J1841-0500 only in the companion paper \citet{mottez_2013_b} (now called paper II), 
in the light of our conclusions about PSR 1931+24.

Here, we examine possible explanations for the behaviour of PSR B1931+24, based, as in \citet{rea_2008}, 
on a putative  companion orbiting the neutron star.
We consider the period of about 35 days mentioned in the observations by \citet{Kramer_2006}.
We furthermore consider two families of hypothesis for the  companions: in this paper
we examine the consequences of a unique companion with an orbital period of 35 or 70 days; a much shorter orbital period is considered in paper II,  
 with a periastron precession period of 70 days.
We begin by briefly reviewing the properties of Alfv\'enic wakes, which result 
from the interaction of orbiting bodies with the magnetospheric or wind plasma, 
and their possible effect on the orbital motion of a  companion.

Table \ref{table_bilan_proprietes} presents
a summary of the observed properties (\textbf{P1} to \textbf{P8}) that eventually are satisfied or not, 
according to the various sets of hypotheses. This could serve as a field guide.


\section{Unipolar inductor and Alfv\'en wings}
\subsection{Alfv\'en wings and associated electric currents}
One of the main questions in connection with  PSR B1931+24 is the interaction of the possible 
companion with the star's environment. The main idea of this paper is based on a system 
of electric currents carried by a pair of stationary Alfv\'en waves, anchored in the pulsar's companion. 

\begin{figure}
\resizebox{\hsize}{!}{\includegraphics{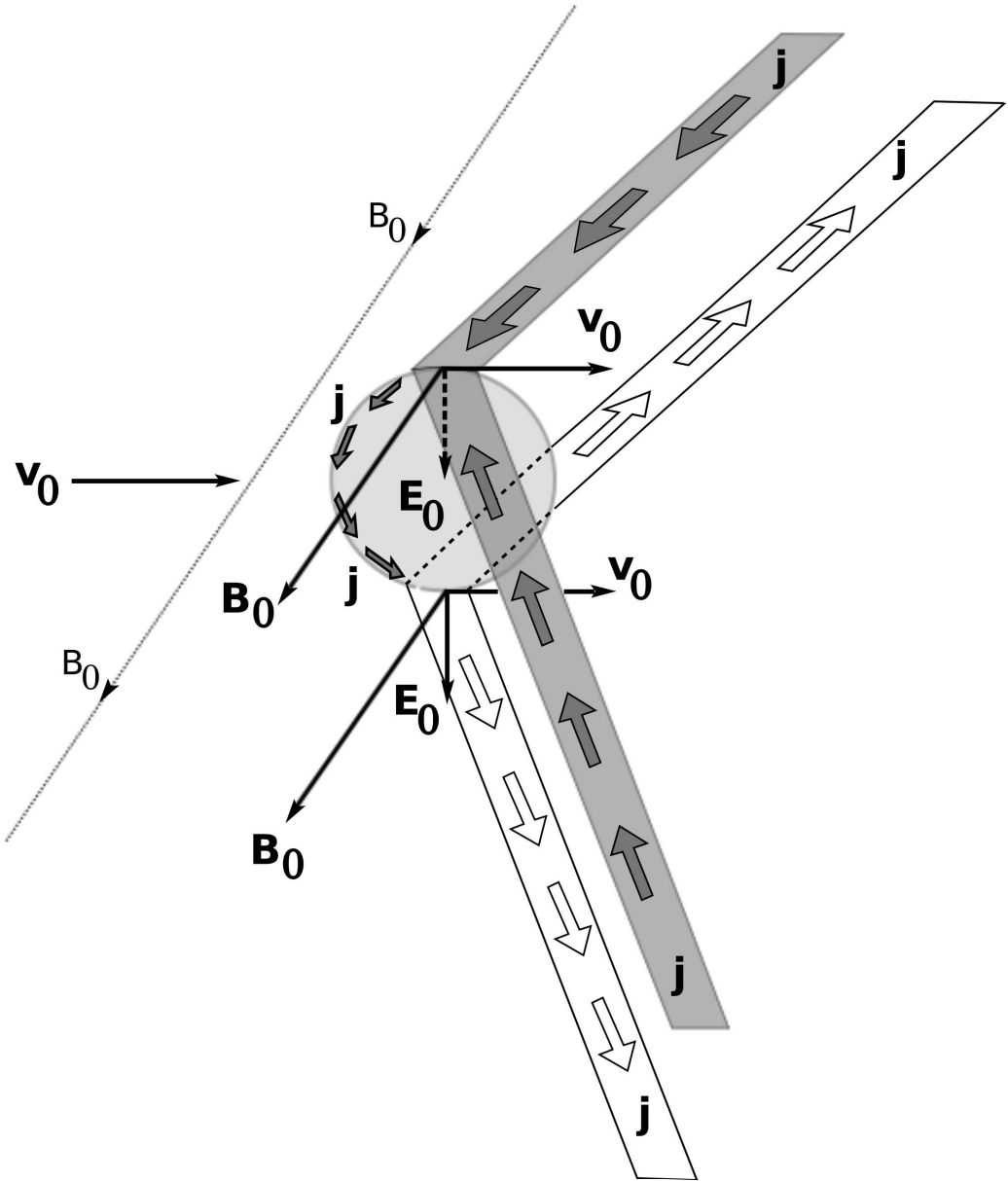}}
\caption{
Schematic view of an unipolar inductor. 
In a pulsar wind, the unperturbed wind's magnetic field $\ve{B}_0$ and  velocity $\ve{v}_0$ are almost, 
but not exactly, perpendicular. The electric field $\ve{E}_0$ created by the unipolar inductor 
is perpendicular to these two vectors; it induces an electric current (of density $\ve{j}$) along the body. 
This current then escapes into the surrounding plasma, 
forming two structures, each of them consisting of an outwards and an inwards current
flow. The current density $\ve{j}$ flowing along the body causes a $\ve{j} \times \ve{B}$ force density
to be exerted on it.}
\label{fig_force_JcrossB}
\end{figure}

In a tenuous and strongly magnetized plasma, the formal Alfv\'en velocity $c_A$
may be higher than the speed of light. The derivation of the linear Alfv\'en wave propagation
velocity $V_A$ must then take into account the displacement current. 
In the plasma frame of reference,
\begin{equation} \label{eq_va}
V_A^{-2}=c^{-2}+c_A^{-2}=c^{-2}+ \mu_0 \rho/B^2,
\end{equation}  
where $\rho$ is the mass density as seen in this frame.
When the wind flows at a velocity $V_0 > V_A$, 
the  companion is enclosed behind a fast-mode bow shock and does not directly interact with the wind. 
But when the pulsar wind flows at a velocity lower than $V_A$,
the  companion is in direct contact with the wind. It then behaves as a unipolar inductor, generating in itself
an electromotive force $\ve{E}_0 = -\ve{V}_0 \times \ve{B}_0$,  where $\ve{B}_0$ 
is the ambient magnetic field and $\ve{V}_0$ is the wind velocity \citep{Goldreich_Lynden_Bell_1969}. 
This generates two systems of current that 
propagate in the surrounding plasma at an angle $\theta$ with the ambient magnetic field, 
forming a so-called Alfv\'en wing. 
\citet{Neubauer_1980} has developed a non-relativistic model of Alfv\'en wings 
to explain the interaction between Jupiter and its satellite Io. \citet{Mottez_2011_AWW} have adapted a linear but relativistic version of that theory. The geometry of an Alfv\'en wing 
is sketched in Fig. \ref{fig_force_JcrossB}. 
The Alfv\'en wing total electric current is  
\begin{equation} \label{neubauer_current}
I_A \sim 4(E_0 -E_i) R_c \Sigma_A,
\end{equation}
where $R_c$ is the  companion radius. 
The factor $\Sigma_A$ is the Alfv\'en wing conductance. 
\citet{Mottez_2011_AWW} have shown that for a highly relativistic Poynting-flux-dominated wind, as expected around pulsars,
\begin{equation}
\Sigma_A \sim \frac{1}{\mu_0 c}.
\end{equation} 
The electric field $E_i$ is caused by the internal resistance of the inductor. In the case of an optimal energy coupling of the  companion and the Alfv\'en wing, as in \citet{Neubauer_1980}
\begin{equation} \label{eq_optimal_coupling}
E_i=E_0/2.
\end{equation}

\subsection{Radio emissions}
\begin{figure}
\resizebox{\hsize}{!}{\includegraphics{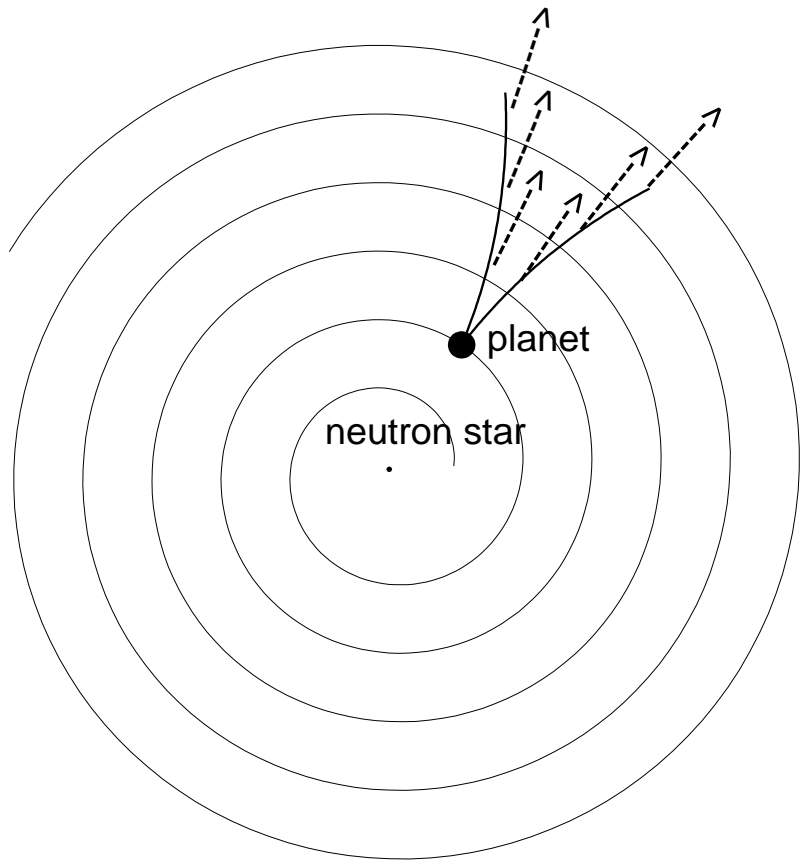}}
\caption{Hypothesis of a planet orbiting in 35 days. Sketch showing the magnetic field line passing through the companion (thin line) and the two Alfv\'en wings (thick lines). The star (thin dot) is at the centre of the figure. The companion is represented by a thicker dot. The arrows indicate the direction of propagation of the radio waves, seen from the observer's reference frame. In the models where $P_{orb} \sim 35$ days, these radiations are emitted from the Alfv\'en wings.}
\label{spirale_AW}
\end{figure}

Alfv\'en wings  are two long ribbons of direct electric currents. In each wing,  part of the current 
is emitted from the  companion and the other part returns to it.
The outward and return currents close 
through the wing-emitting body.
The distance between the two current lines in one wing is of the order of $R_c$, the companion's radius.
The current that circulates along a wing 
is related to the Alfv\'en wave conductivity $\Sigma_A$ \citep{Neubauer_1980}. 
If there is a stationary flow, the Alfv\'en wings carry a DC current.
However, plasma and current instabilities may add an AC component to it. In that case,
Alfv\'en wings would behave like two large electromagnetic antennas. 
For instance, the Alfv\'en wings associated to the Io-Jupiter flux tube are 
the sources of strong radio emissions \citep{Queinnec_1998}. Small-scale Alfv\'en waves are trapped in the wings,
which cause particle acceleration \citep{Su_2003, Ergun_2006} and decametric radiation 
\citep{mottez_2007_b,mottez_2008_b,mottez_2009_b}. In the highly relativistic pulsar wind, 
the acceleration and emission process may be quite different from those arising in the vicinity of Jupiter, 
but the current $I_A$ certainly constitutes an important source of free energy for radio emissions. 
\citet{Mottez_2011_AWW} have suggested that the strong current in the pulsar companion's Alfv\'en wings 
could be unstable and might cause powerful radio emissions. The generation mechanism of 
these emissions has not been studied yet, but preliminary results are presented in \citet{Mottez_2011_Graz},
assuming that the radio-source is convected with the pulsar's wind
when it passes the Alfv\'en wing. 
In that case, because of relativistic aberration and whatever the emission directions  of the radio waves in the the wind frame, they would make
a small angle $\theta_\infty$ in the observer's frame   with the direction of the wind velocity given by
\begin{equation} \label{eq_theta_infty}
\theta_\infty \sim {\gamma_\infty^{-1}},
\end{equation}
where $\gamma_\infty$ is the asymptotic value of the wind's Lorentz factor.
The observer's frame is that of an observer on Earth. It is roughly the same as the 
frame defined by the motion of the pulsar's companion.
Fig \ref{spirale_AW} is a sketch showing the magnetic field line passing through the pulsar companion. 
At this distance, it has almost the shape of an Archimedes spiral. The planet is represented by a black dot. Two thick lines emerging from it show the global shape of the two Alfv\'en wings, making a constant angle with the ambient magnetic field. The arrows show the direction  (as seen in the observer's frame) of the radio waves emitted by the pulsar wind when it crosses the Alfv\'en wing.  
The Lorentz factor $\gamma_\infty$, though unknown, 
is expected to be large  \citep{Michel_1969,Henriksen_1971,Contopoulos_1999,Michel_2005,Bucciantini_2006}, therefore
the cone angle of the radio emission is very narrow and the radio emission associated to the Alfv\'en wing 
would be observable only during a brief lapse of time. 

Alternatively, the source might be firmly connected to the planet. For instance, 
radio waves could be emitted from its (potentially present) ionosphere when it is disturbed 
by fluctuations of the current flowing along it. In that case, the radio source would not be fast in 
the observer's frame and would not be subject to any relativistic aberration.
Until a precise mechanism of radio emission is elaborated, it will not be possible to make 
any prediction concerning its directivity.  Therefore, we assume 
(at least in some of the models) that the planet and / or the Alfv\'en wing is a radio source, 
but we make no \textit{a priori} assumption on the directivity of their emission. 

The assumption about the radio-waves from Alfv\'en wings does not imply that the radio waves from standard pulsars emanate from Alfv\'en wings. We follow the general proposition that pulsar radio emissions are emitted from well inside the light cylinder, and that the origin of these waves have no connection with Alf\'en wings. But in some of the models tested in this paper for extremely intermittent pulsars, we simply consider that the radio waves from Alfv\'en wings would \textit{mimic} standard pulsar radio emissions, and this would happen only for the very small and very peculiar class of pulsars represented by PSR1931+24. Moreover, this is an hypothesis that we decide to consider, and not an assertion that we necessarily defend until the end of the paper.

\subsection{Force exerted by the current system onto the  companion}
Figure \ref{fig_force_JcrossB} sketches the configuration of the current system  in the vicinity of the  companion. 
This figure represents
the  companion  in the pulsar wind, 
at a large distance form the light cylinder. The current closing the Alfv\'en wing circuit 
through the companion shown in this figure
generates a force density $\ve{J} \times \ve{B}$
acting onto the  companion or its immediate environment by a magnetic thrust \citep{Mottez_2011_AWO}. 
Whether this force is mainly radial or azimuthal depends on the orientation of the wind velocity and the magnetic field. 
The Lorentz force density can be integrated along the current flowing on the companion's surface. Roughly estimated, 
the modulus of the total force exerted onto the  companion is
\begin{equation} \label{eq_force_magnetique}
F=2 R_c I B_F, 
\end{equation}
where $B_F$ is the component of the magnetic field that is perpendicular to the plasma flow and
$I$ the total current flowing in the Alfv\'en wings.
As long as these
wings exist, this force is continuously exerted and may ultimately
influence the evolution of the companion's orbit. 
In our other estimates, we denote with $F_r$ the radial components of $\ve F$ and with $F_t$ its ortho-radial component 
lying in the orbital plane of the  companion, whose mass is $M_c$. The pulsar's spin frequency is $\Omega_*$.
\citet{Mottez_2011_AWO} have calculated that
\begin{equation}  \label{FtetD_d} 
\frac{F_r}{M_c}=\frac{C}{r^2} \quad \mbox{ and } \quad \frac{F_t}{M_c} = \frac{D}{r^3},
\end{equation}
where $C$ and $D$ are constant factors,
\begin{equation} 
C =\frac{4 R_c^2 \Omega_*^2 \Phi^2}{\mu_0 c^2 M_c}, \qquad \qquad  
D = \frac{4 R_c^2 \Omega_* \Phi^2}{\mu_0 c M_c},
\label{def_D}
\end{equation}
and
\begin{equation} \label{eq_vent_Phi}
\Phi = r^2 B_0^r
\end{equation}
is the magnetic flux, which is a constant of motion along a streamline \citep{Beskin_1998,Bucciantini_2006}.
A Keplerian orbit is represented in polar coordinates in the orbit's plane by the equation
\begin{equation}
r = \frac{a \, (1 - e^2)  }{1 + e\, \cos \phi}.
\label{eqtrajectoire}
\end{equation}
The average changes of $a$ and  $e$ over an orbital period \citep{Mottez_2011_AWO} are
\begin{eqnarray} \label{eq_delta_a}
< \frac{da}{dt}> &=& \frac{\Delta a }{P} =  2 a \, \frac{D}{\sqrt{GM_* a^5}} \ \, \left(\frac{2 + e^2}{2(1 - e^2)^{2}}\right)
\\ \label{eq_delta_e}
< \frac{de}{dt}> &=& \frac{\Delta e}{P} =  \frac{3}{2} \, \frac{D}{\sqrt{GM_* a^5}} \ \, \frac{e}{(1- e^2)}.
\end{eqnarray}
These variation rates both have the same sign as $D$, which is fixed by that of $\Omega_*$.
For a prograde orbit, $D>0$, $a$ and $e$ increase, the orbit becoming more eccentric and distant. 
Therefore, the Alfv\'en wing thrust tends to chase the body away from the star. A retrograd orbit 
evolves towards a circular shape with a decreasing semi-major axis.

\subsection{Possible location of the fast-mode critical surface}

All these considerations are valid if the planet is immersed in a sub-Alfv\'enic pulsar wind, by which we mean that 
the modulus of the relative velocity between the companion and the pulsar's wind is
lower than the modulus of the relativistic Alfv\'en speed $V_A$. For a non-relativistic orbital velocity of the companion, 
this condition essentially constrains the wind's speed.
It is therefore important to know if the pulsar wind can make a transition from a sub-Alfv\'enic to a super-Alfv\'enic 
velocity, and to know the distance to the star of this so-called fast-mode critical surface. 
Many pulsar wind models have been developed, one of the simplest is that by  \citet{Michel_1969}, which
accounts for the magnetic field, the star's rotation, and the wind particles masses, but neglects 
the dipole inclination and the gravitation. The model considers the poloidal field $B_r$ to be 
radial and includes the self-consistent toroidal field $B_\phi$. 
Other models for steady-state and axisymmetric, but not necessarily radial, winds have
been developed more recently \citep{Beskin_1998,Bucciantini_2006}. 
Their equations admit a set of integrals of motion along stream lines, in particular,
the magnetic flux $\Phi$ defined in equation (\ref{eq_vent_Phi}) and the mass flow  $f$
\begin{equation} \label{eq_vent_f}
f=\rho v_r r^2, 
\end{equation}
where $\rho$ is the mass density and $v_r$ the wind's radial velocity.
The magnetization parameter $\sigma_0$ of the pulsar wind, 
\begin{equation} \label{eq_sigma_0}
\sigma_0=\frac{\Omega^2 \Phi^2}{\mu_0 f c^2},
\end{equation}
is also constant along a streamline.
\citet{Mottez_2011_AWW} have shown than in cold ideal magneto-hydrodynamic Poynting-flux-dominated winds, where 
the models show that asymptotically $\gamma_\infty \sim \sigma_0^{1/3}$, 
the wind is sub-Aflv\'enic, in the sense defined above, at any distance. But if $\gamma_\infty \sim \sigma_0^{a}$, and $a> 1/3$, 
a transition from sub-Alfv\'enic to super-Alfv\'enic wind occurs at a finite distance. 

Observations of the equatorial sectors of winds in 
pulsar wind nebulae show lower values of the 
asymptotic magnetization \citep{Kennel_1984a,Kennel_1984b,Gaensler_2002}, and some authors 
have suggested that the asymptotic value of the Lorentz factor is $\gamma_{\infty} \sim \sigma_0$. 
\citet{Arons_2004} has argued that
to understand the observed high Lorentz factors and the low magnetisation,
dissipation of magnetic energy probably occurs  in the asymptotic wind zone.
\cite{Begelman_1994} have shown that 
when  flux tubes diverge faster than radially, the fast magnetosonic point can occur 
closer to the light cylinder. 
Then, the existence of Alfv\'en wings would depend on the distance from the star to the planet 
and would be a result of
a non-radially-diverging wind flow and / or dissipation. 
According to \citet{Michel_2005}, the finite angle between the pulsar rotation axis and the magnetic axis makes the standard 
axisymmetric models mostly inappropriate. 
It therefore cannot be taken for granted that 
the fast-mode critical surface is rejected to infinity. The latter is
defined for a cold wind 
by $v_r = V_A$, where $v_r$ is the radial wind's speed
and $V_A$ the relativistic Alfv\'en speed (Eq. (\ref{eq_va})) calculated from the modulus of the field.
It is possible that the companion spends only a finite part of its orbit in 
sub-Alfv\'enic wind regions (where $v_r< V_A$), or even that it never finds itself in a sub-Alfv\'enic wind. 
In the latter case our suggestions would be pointless.


\section{Companion with an orbital period of 35 days?} \label{orbite_35_jours}

Our first hypothesis is that a companion orbits the star with a 35-day period. 
The size of 
the orbit is $a_R \sim 0.23$ AU $\sim 900 c/\Omega$, where $c/\Omega$ is the light cylinder radius. 
The orbit can sometimes be in a super-Alfv\'enic wind, and sometimes in a sub-Alfv\'enic wind. 
When in a sub-Alfv\'enic wind, two Alfv\'en wings are formed, and we assume
that they emit  radio waves. We examined whether these waves could be the signal 
that is observed when the pulsar is on.
When the companion is in a super-Alfv\'enic wind, there would be 
no Alfv\'en wings and the pulsar would be off.

This hypothesis suggests that we would not see the actual radio emissions from the extremely intermittent pulsar itself (maybe because we are in the wrong direction to see them), but only radio waves emanating from the Alfv\'en wings of their companion.

\subsection{Electric current carried by the Alfv\'en wing}

\begin{figure}
\resizebox{\hsize}{!}{\includegraphics{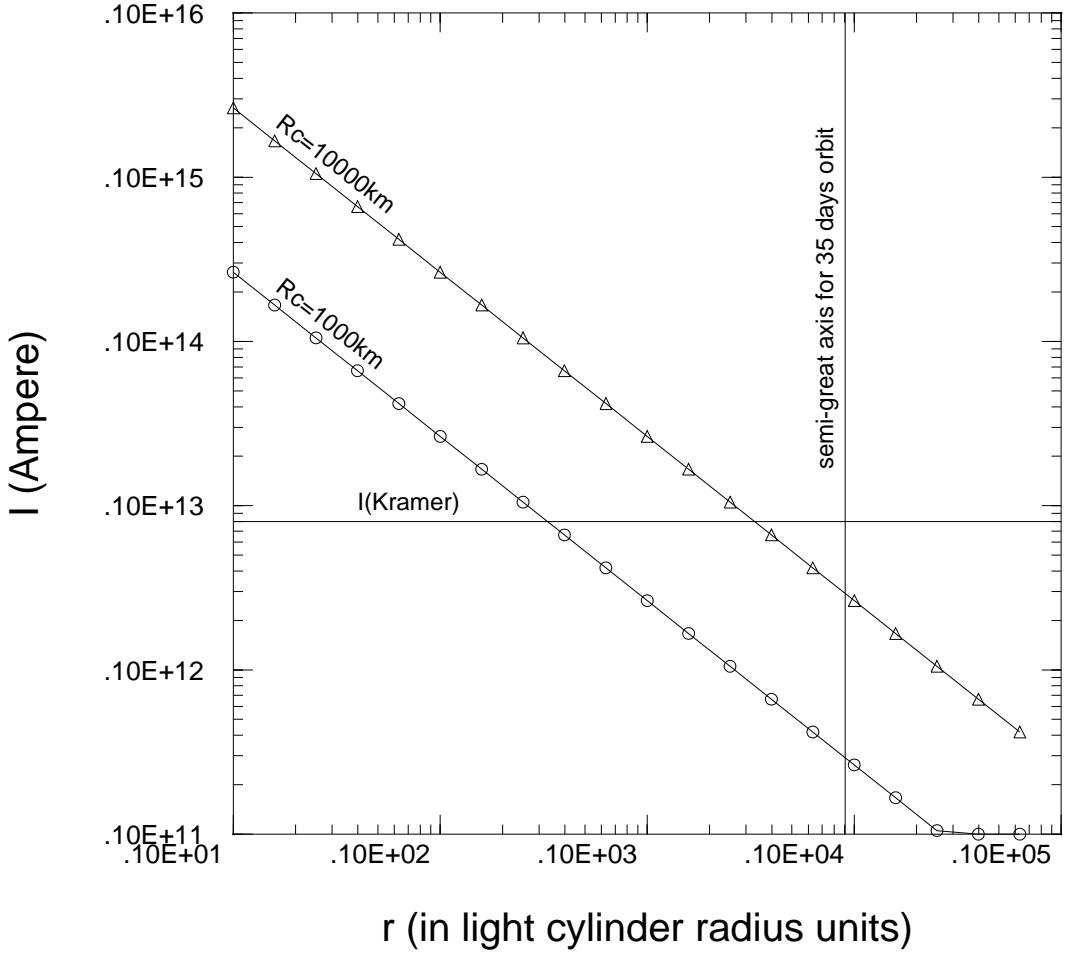}}
\caption{Hypothesis of a planet orbiting in 35 days. The dependency of the electric current carried by the Alfv\'en wings as a function of the distance. Its algebraic expression is given by Eq. (\ref{eq_total_current}), with $E_i={\Omega_* \Phi}/{2r}$ (case of an optimal electric coupling). The distances are given in units of the light cylinder radius $R_{lc}$. The distance corresponding to the semi-major axis is indicated by the vertical line. For comparison, the current computed in \citep{Kramer_2006}, given in Eq. (\ref{additional_torque_kramer}) is indicated by the horizontal line.}
\label{courants_AW}
\end{figure}

The lack of knowledge of the wind velocity 
does not preclude us from estimating the current carried by the Alfv\'en wing when the  companion is in the sub-Alfv\'enic regime. Taking equations (\ref{neubauer_current}) and (\ref{eq_optimal_coupling}) into account,
the definition of the induced electric field $E_0$ and the fact that at a large 
distance from the light cylinder $v_\phi<< \Omega r$, the electric current carried by 
an Alfv\'en wing is, for the best energy coupling of the planet and the Alfv\'en wing
\citep{Mottez_2011_AWW}
\begin{equation} \label{eq_total_current}
I_A =  2 \frac{\Omega_* \Phi}{r}  R_c \Sigma_A.
\end{equation}
When the energy coupling of the companion to the Alfv\'en wing is
less than optimal,  
equation (\ref{eq_optimal_coupling}) overestimates the current $I_A$.
In Fig. \ref{courants_AW}, the Alfv\'en wing currents deduced from 
equation (\ref{eq_total_current}) are plotted as a function of the radial distance between the star and its companion 
for two different values of the companion's radius. 
For a 10 000-km-wide planet, it is possible to reach a current weaker than, but comparable to,  $\Delta I_{pc}$ given by Eq. (\ref{additional_torque_kramer}), i.e. the  
Goldreich-Julian current expected from the polar cap of PSR B1931+24. 
The Goldreich-Julian current is widely  believed to be the source of the standard pulsar radio emissions.
Accordingly, the Alfv\'en wing current associated to a 10 000-km-wide planet could be a radio source of comparable intensity.


\subsection{What would we observe with radio telescopes?}

Most of the theories of PSR B1931+24 (e.g. \citep{rea_2008,Cordes_2008}) consider that the off state  
results from an alteration of the polar cap current of the star. The fact that  $\Delta I_{pc}$ of 
Eq. (\ref{additional_torque_kramer}) associated to the variation between $\dot P_{on}$ and $\dot P_{off}$ \citep{Kramer_2006} 
is of the order of the expected current from the polar cap of the star reinforces this way of considering the problem. 
But at the same time, an Alfv\'en wing starting from a  companion with a 35-day period, and therefore far beyond 
the light cylinder of the pulsar, could not reach the star because the radial wind velocity
exceeds the velocity of radially propagating shear Alfv\'en waves from the 
Alfv\'en radius on. This radius is slightly smaller than the light cylinder radius. 
An object orbiting the star at larger distances could then act on the star's surface or on 
its inner magnetosphere only by dropping matter on it
or by emitting fast-mode signals that could reach it.
The first 
possibility has been explored by \citet{Cordes_2008}. 
Although fast magnetosonic waves could propagate down to the star
provided they are emitted below the fast-mode critical surface, the isotropy of their propagation properties 
imply that the energy carried by such signals declines 
as the inverse square of the distance travelled. Therefore, only 
a negligible fraction of the energy emitted in this form at a large distance 
would eventually reach the star.

Since we examine here only electromagnetic types of coupling, we conclude that an object orbiting at a
much larger distance than the light cylinder radius $R_{lc}$ would be unable to affect the emission process
taking place closer than this distance from the star, and in particular, at the pulsar's polar cap. This means that 
if the observed transient radio emissions are to result from the interaction of the companion 
with the pulsar wind, their source should be
in the vicinity of the  companion itself. This is not impossible, because as we noted above, the current in the Alfv\'enic wake could  be similar to the Goldreich-Julian current.

Because radio emissions are seen during several days without interruption there must be a large set of emission directions. 
Therefore, the radio source is probably not conveyed at the wind's velocity.
Instead it is probably connected to the planet, possibly to its ionosphere.  
 
However, since the radiation is pulsating, these emissions are very likely still endowed with some directionality,
so that  the varying direction of the local magnetic field could result in some modulation of the radiation detected at Earth.
The direction of the field would vary because of  
the inclination of the pulsar's magnetic moment with respect to the rotation axis and oscillate around the neutron star equator.
The periodicity of this modulation 
is the pulsar spin period. 
The observed frequency $\Omega_{obs}$ of the  signal received by the observer
is the difference or the sum of the pulsar's rotation frequency $\Omega_*$ and the orbital frequency of the companion $\Omega_{orb}$:
\begin{equation} \label{eq_composition_frequences}
\Omega_{obs} = \Omega_* \pm \Omega_{orb}.
\end{equation}
It is a sum if the companion orbit is retrograde, and a difference if it is prograde.
The wavelength of the modulation 
is $\lambda_{mod} =V_a P_{obs} \sim 2 \pi c/\Omega_*$. 

The fact that we detect the radiation emitted by the  companion does not mean that the pulsar does not radiate. 
Indeed, the position of PSR B1931+24 in the $P-{\dot P}$ diagramm, far from the death line and typical 
of most pulsars, strongly suggests that the pulsar is active, and "normal". 
But we could simply be situated out of its emission beam. Otherwise, two kinds of emissions would be seen,
those from the neutron star (possibly its polar cap),
which would be permanently active, 
and those associated to the  companion and the Alfv\'en wings, which would only be observable
when the pulsar is on.


\subsection{Does the orbit cross the sub to super-Alfv\'enic frontier in the pulsar wind?}

Now, it should be explained why the radio emissions emanating from the two  companion's Alfv\'en wings sometimes stop.
The on and off states may be simply arise because sometimes the  companion is in a sub-Alfv\'enic wind, sometimes in a super-Alfvenic wind. When in the sub-Alfv\'enic wind, Alfv\'en wings would be present, with
subsequent radio emissions. When in the super-Alfv\'enic wind, the Alfv\'en-wings do not exist because
the companion is separated from the wind by a shock front instead. The shutting down
of the associated extended current branches causes the radio emissions to cease. 
This hypothesis can explain why the pulsar radio emissions stop so abrubtly: 
when the  companion is at the limit of the sub-Alfv\'enic domain, 
the plasma flow is very fast, the magnetic field is still strong, and $I_A$ shows no sign of progressive extinction, in accordance with property P6.

The hypothesis of a transition from a sub- to a super-Alfv\'enic regime also explains why the transitions 
between the "silent" and "active" states, in spite of being quasi-periodic, are nevertheless unpredictable: 
because the position of the frontier between the two wind regimes is expected to be highly fluctuating around an average value, as is the case of almost all frontiers in the magnetospheres explored up to now.


\subsection{Circular or an elliptical orbit in the star's equatorial plane?}

We first consider a  companion orbiting near the equatorial plane of its star. 
If its orbit is circular, then the orbital radius should correspond by chance
to the region close to the sub- to super-Alfv\'enic transition. 
The changes from the off state to the on state would then not be periodic, 
but would depend on the fluctuations of the pulsar wind. Because there is no reason for a quasi-periodicity 
of 35 days in these fluctuations, there should be no
quasi-perodicity in the pulsar mode transitions, which contradicts property P2.
This is why we do not retain the hypothesis of an equatorial circular orbit. 
If the orbit is in the star's equatorial plane, it must be elliptic and the $\sim$ 25 days of off 
mode correspond to the time spent near the apoastron, in the super-Alfv\'enic wind, 
the rest of the time corresponding to closer distances in the sub-Alfv\'enic wind. As the 
time spent in the on state ($\sim$ 5-10 days) and in the off state ($\sim$ 20-25 days) is well marked, 
the orbit ellipticity probably is not negligible.

This explanation nevertheless has a serious drawback: 
if the  companion is in an elliptical orbit, its distance from the star varies and thus also
the distance from the star to the radio source, whether it is attached to the companion
or to the Alfv\'en wings. 
There must be a  Doppler shift associated to this varying distance that induces a modulation 
of the pulsar's observed period $P$, which contradicts the property P7.  
Furthermore, the time-varying orbital velocity of the companion is an another cause of variation of $P$.
We consider the variation of the orbital 
frequency, i.e. the azimuthal frequency (equation (\ref{eq_composition_frequences})). 
The angular momentum conservation
implies that the areal velocity $H=\Omega(r) r^2/2$ is a constant of the motion and 
$H=\Omega_{orb} a^2 (1-e^2)^{1/2}/2$. Let $\lambda = r_{min}/a$ be the ratio of the periastron to the semi-major axis, 
and $k= \Omega_{orb}/\Omega(r_{min})$ the ratio of the mean orbital frequency to the orbital angular velocity at the periastron.
Some algebra shows that 
\begin{equation} \label{eq_lambda}
k^2 = {\lambda^3}/{(2 - \lambda)}.
\end{equation}
For a circular orbit, the solution is obviously $\lambda=1$ for $k =1$. For a weakly 
elliptic orbit ($\lambda$ and $k$ close to unity), the solution is obtained by differentiation:
\begin{equation} \label{eq_rmin}
\frac{r_{min}}{a}=1 - \frac{1}{4}\left(1 - \frac{\Omega_{orb}}{\Omega(r_{min})}\right).
\end{equation}
This relation combined with Eq. (\ref{eq_composition_frequences})  
shows that the circularity of the orbit is strictly constrained by the invariance of $P=2 \pi/\Omega(r)$ (condition P7). 
Numerically, a ratio ${r_{min}}/{a}=0.99$ implies 
${\Omega_{orb}}/{\Omega(r_{min})}=0.96$. In that case, there should be a periodic variation  
$\Delta P= -P \Delta \Omega/\Omega = 3.2 \times 10^{-2}$s
that is not observed. Thus  we must reject the hypothesis of an elliptical equatorial orbit.

\subsection{Inclined circular orbit?}

\begin{figure}
\resizebox{\hsize}{!}{\includegraphics{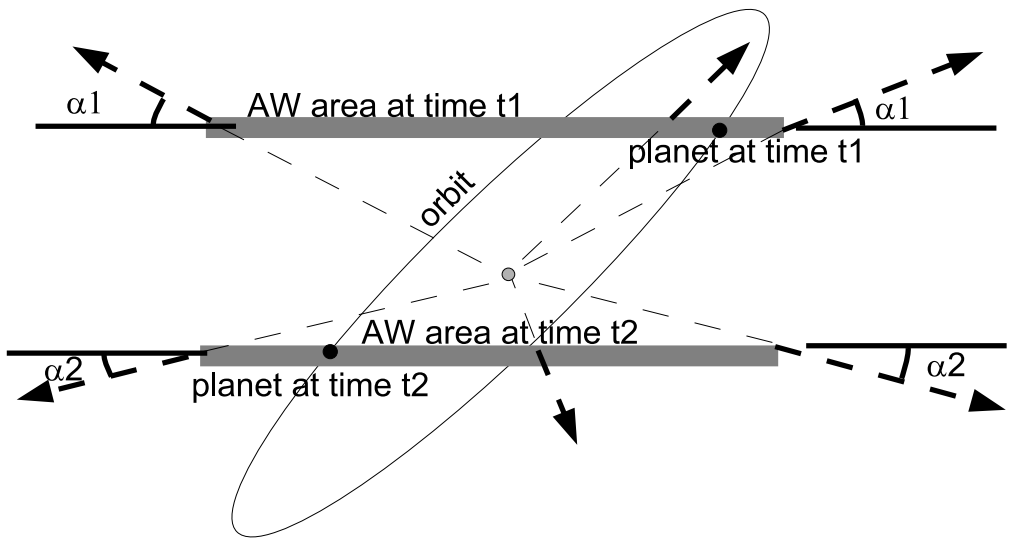}}
\caption{Hypothesis of an inclined circular orbit. Sketch showing the orbit and the planet at two different times $t_1$ and $t_2$. The shaded area corresponds roughly to the area where the Alfv\'en wing is situated. The dotted lines with an arrow are the directions of the radio emissions. They are emitted radially, from the Alfv\'en wings. The angle $\alpha$ that is made between the radial direction and the plane defining the position of the Alfv\'en wings determines the angle of view, i.e. the star's equatorial plane. If the observer's angle of view relative to the star's equatorial plane is also $\alpha$, he can see the radiation emitted from the Alfv\'en wings. This angle is not time invariant, it depends on the position of the planet on its orbit.}
\label{orbite_inclinee}
\end{figure}

If the orbit is circular and inclined upon the star's equatorial plane, the  companion may
be close to the equatorial plane (when it is near the nodes of its orbit in the star's equatorial plane)
or far from this plane. According to 2D pulsar wind models, such as those of \citet{Bucciantini_2006}, 
the wind's velocity (or Lorentz factor) depends on the latitude. We assume that as for the low $\sigma_0$ 
that these authors have simulated, the wind is faster near the equatorial plane than far from it, and we consider
a  70-day orbit. We assume that the companion
travels in a super-Alfv\'enic wind (for most of the time) near each node, then is in the off mode, 
and spends about five days at higher latitudes, and five days at lower negative latitudes in a sub-Alfv\'enic wind, 
then is in the on mode.  
This is consistent with property P2.  
If the radio source resides in wind material, the radio waves would be emitted almost radially,
in a direction that differs according to whether  
the companion is at its highest latitude or nearer the equatorial plane.
This is shown in Fig. \ref{orbite_inclinee}.
It may happen that in the on mode the  companion is at a favourable latitude to being seen from Earth. 
Later, the Earth would be out of its emission beam 
and radio emissions would not be seen. This may explain
the existence of an off and an on regime (property P1). But the companion should also 
only be on when it is in a particular physical regime, such that the rate of damping 
$\dot P_{on} \ne \dot P_{off}$ (property P3).
To adopt the hypothesis of an inclined circular orbit in a latitude-dependent wind, 
we would have to admit that the on regime simultaneously corresponds to a favourable 
line of sight and to a favourable physical regime. 
This coincidence seems too strong, which
is an encouragement to reject the hypothesis of a circular inclined orbit. 

\subsection{Could the difference in period drift result from orbital evolution driven by the Alfv\'en drag?}\label{sec_calcul_p_dot}

When the companion is in the sub-Alfv\'enic wind and Alfv\'en wings are present,
the associated electric current combined with the ambient magnetic field exerts a force that
acts on the companion's orbit and causes a secular change $\dot P_{orb}$ of the mean orbital period.
{We now discuss whether}  this phenomenon can explain the difference of the observed period time derivatives
$\dot P_{on} > \dot P_{off}$ (property P3). 

If the difference of period derivatives in the on an the off regimes were 
to be caused by a non-zero derivative of the orbital period during the on regime,
this difference could be found 
from Eq. (\ref{eq_composition_frequences}), since
\begin{displaymath}
\dot \Omega_{off} = \dot \Omega_* \mbox{ and } \dot \Omega_{on} = \dot \Omega_* \pm \dot \Omega_{orb}.
\end{displaymath}
Substracting these two equalities, and expressing $\Omega_{orb}$ as a function of $P_{orb}$,
\begin{equation} \label{eq_dot_P_orb_observations}
\dot P_{orb}= \dot P_{orb,on} =\pm  \frac{P_{orb}^2}{2 \pi}(\dot \Omega_{off} -\dot \Omega_{on}).
\end{equation}
For $P_{orb} \sim$ 35 days and the known values of $\dot \Omega_{off}$ and $\dot \Omega_{on}$, 
$\dot P_{orb}=-5 \times 10^{-2}$. 

Combining Eqs. (\ref{def_D}) and (\ref{eq_delta_a}),
the electromagnetic forces should give rise to a 
variation of the semi-major axis per orbit given by 
\begin{equation} \label{eq_Delta_orb_a_0}
\Delta_{orb} a = \frac{4 R_c^2 \Omega_* \Phi^2 P_{orb}^2 }{a^3 \pi \mu_0 M_c}\, \left(\frac{2 + e^2}{2(1 - e^2)^{2}}\right). 
\end{equation} 
This can be transformed into a variation of orbital period by log-differentiating Kepler's third law:
\begin{equation} \label{eq_dot_P_orb_theorie}
\dot P_{orb, theory} =\Delta_{orb} P_{orb}/P_{orb} = {3 \Delta_{orb} a}/{2 a}.
\end{equation}
The values of $\dot P_{orb}$ deduced from the observations, as given by  Eq. (\ref{eq_dot_P_orb_observations}),
should be consistent with the theoretical estimate $\dot P_{orb, theory}$ in Eq. (\ref{eq_dot_P_orb_theorie}).
We have considered the case of an Earth-like planet, a 100-km-wide body and an 1-km-wide asteroid, with a circular orbit.
The results are presented in Table \ref{table_bilan_35days}. In this table, the current $I_A$ is derived from Eq. (\ref{eq_total_current}).
It can be seen that the secular orbital drift induced by the Alfv\'en wing force is too weak by many orders of magnitude. 
Any reasonable value of the eccentricity cannot compensate this large difference. This is one more reason to 
reject the hypothesis of a companion with a 35- (or 70-) day orbit: 
the difference between the period derivatives $\dot P_{on}$ and $\dot P_{off}$ cannot be quantitatively explained. 
In other words, it does not fit with property \textbf{P3}.

\begin{table*}
\caption{Effect of the Alfv\'en wing force estimated for various bodies orbiting the pulsar in 35 days.} 
\label{table_bilan_35days} 
\centering 
\begin{tabular}{|c | c c c c c c c c|} 
\hline\hline 
  & diameter  &volumic mass  & $I_A$ &$F_{AW}^r$   &$F_{AW}^t$ &$\Delta_{orb} a$ &$\dot P_{orb, theory}$  & $\dot P_{orb, theory}/\dot P_{orb}$\\ 
  & (km)  &(kg.m$^{-3}$)  & (A) & (N)   &(N) &(m, per orbit) & (dimensionless)  & (dimensionless)\\ 
\hline \hline
planet & 10 000 & 5000 & $1.5\times 10^{11}$ & $3. \times 10^{16}$ & $3.2\times 10^{13}$ & 8,9 & $3,8\times 10^{-10}$ & $\sim 10^{-8}$ \\
\hline
small body & 100  km & 3000 & $1.5\times 10^{9}$ &$2.9 \times 10^{12}$ & $3.2\times 10^{9}$ & 1563 & $6.7\times 10^{-8}$ & $\sim 10^{-6}$ \\
\hline
asteroid & 1  km & 3000 & $1.5\times 10^{7}$ &$2.9 \times 10^{8}$ & $3.2\times 10^{5}$ & 156316 & $6.7\times 10^{-6}$ & $\sim 10^{-4}$ \\
\hline \hline

\end{tabular}
\end{table*}


\section{Companion at short distance and a periastron precession with a period of $\sim$ 70 days?} \label{periastre_70_jours}

We have now accumulated enough arguments 
to reject the hypothesis of a 35- or 70-day orbital period and, together with it,
all the inferences developed in section \ref{orbite_35_jours} (except for the rejection of the hypothesis that supports them).

A periodicity of 35 days could nevertheless be obtained if we assumed 
a short orbital period, the periodicity of 35 days being not the orbital period, but the period of the precession of the periapsis
of the orbit.
This is possible because the pulsar is a compact star.
More exactly, as shown in paper II,
the period of the precession of the periapsis should be twice as long, i.e. $P_{per} \sim 70$ days.
The rotation angle $\Delta \varphi_{per}$
of the periapsis per orbit, derived from the theory of general relativity,
\begin{equation} \label{eq_gr_precession}
\Delta \varphi_{per} = \frac{6 \pi G M_* }{ c^2 a (1-e^2)},
\end{equation}
and the period of the precession of the periapsis then is 
\begin{equation} \label{eq_gr_period_precession}
P_{per} = \frac{2 \pi P_{orb}}{\Delta \varphi_{per} }=
\frac{2 \pi c^2 a^{5/2} (1-e^2)}{3 (G M_*)^{3/2}}.
\end{equation}
Asteroids originally orbiting along a common orbit keep on sharing the same orbit at later times, because
the precession rate of each one's orbit is the same.
The idea that PSRB 1931+24 could be orbited by a close swarm of asteroids is discussed in  paper II.


\section{Conclusions}

Our analysis was based on the series of observed peculiarities \textbf{P1}--\textbf{P8}
of the pulsar PSR B1931+24. 
We tried to identify a theoretical explanation that would be consistent with all these properties. 
We assumed that the quasi-periodicity of 35 days of the behaviour of PSR 1934+21 (property \textbf{P2}) 
results from a single body orbiting the neutron star at this period, or twice this period. We considered
that the coupling between the planet, the star, and the radio emissions is caused by 
the Alfv\'en wings carried by the planet when it moves in a sub-Alfv\'enic plasma (the pulsar's wind, or its magnetosphere).
These models implied that the radio waves received from PSR 1934+21 were not the actual pulsar radio waves (emitted from inside the light cylinder), but waves attached to the pulsar's companion. Therefore, these waves were supposed to mimic those of a normal pulsar. This might seem improbable and we could have discussed that point. But we did not because the models could be rejected on other bases that we judged better suited for quantitative investigation.
No model based on the assumption of a body orbiting in 35 days or 70 days  
could be made to be consistent with all properties \textbf{P1}--\textbf{P8} however. 

Nevertheless, the quasi-periodic succession of on and off phases might be induced
by a companion, if we consider that 70 days is not its orbital period, but the 
\textit{period of the precession of the periastron}. This precession would 
be induced by the relativistic gravitation field of the neutron star.  
This implies that the body would be orbiting very close to and possibly inside the light cylinder.
Under these conditions, a large companion would be disrupted by tidal forces. A stream of smaller bodies, 
possibly asteroids, would result from this disruption. 
A direct electromagnetic interaction between the star and its companions could then be considered.  
A more detailed analysis of this idea is developed in the companion paper II.

\newpage

\begin{table*}
\caption{Summary of the different hypotheses and of the properties they satisfy, or fail to satisfy. } 
\label{table_bilan_proprietes} 
\centering 
\begin{tabular}{|c | c c c c c c c c|} 
\hline\hline 
  & {\bf P1}  &{\bf P2}  &{\bf P3}   &{\bf P4} &{\bf P5} &{\bf P6} &{\bf P7} & {\bf P8}\\ 
  & on/off & $\sim$ periodic & $\dot P_{off}<\dot P_{on}$ & standard & equivalent& fast & $P_{on}=P_{off}$ & no timing \\ 
  &        &                 &                            & $P \dot P$& $\Delta I_{pc}$& transition &          & residual\\ 
\hline \hline
neutral bodies & yes & no & yes & yes & yes & --& yes& yes\\
\hline \hline
$P_{orb} \sim 35$ or $70$ days. &&&&&&&&\\
\hline 
circular equatorial  & yes &  no & wrong values & yes & OK... & yes & yes& yes\\
\hline 
elliptical  & yes &  yes&  wrong values & yes & ... for Earth-like & yes & no& yes\\
\hline
circular inclined  & yes & yes & wrong  values & yes & planet only. & yes & no& yes \\
\hline \hline

\end{tabular}
\end{table*}

\end{document}